\documentstyle[epsfig,psfig]{texas}

\begin{document}

\title{Hydrodynamics of black hole--neutron star coalescence.}
\author{William H. Lee$^{1}$ and W\l odzimierz Klu\'{z}niak$^{2,3}$}
\address{(1) Instituto de Astronom\'{i}a--UNAM\\
Apdo. Postal 70--264\\
Cd. Universitaria\\
M\'{e}xico D.F. 04510\\
MEXICO\\
}
\address{(2) University of Wisconsin--Madison\\
Physics Department\\
1150 University Ave.\\
Madison, WI, 53705\\
}
\address{(3) Copernicus Astronomical Centre\\
ul. Bartycka 18\\
00--716 Warszawa, Poland\\
{\rm Email: wlee@astroscu.unam.mx, wlodek@astrog.physics.wisc.edu}}

\begin{abstract}
We present a numerical study of the hydrodynamics in the final stages
of inspiral in a black hole--neutron star binary, when the separation
becomes comparable to the stellar radius. We use a Newtonian
three--dimensional Smooth Particle Hydrodynamics (SPH) code, and model
the neutron star with a soft ($\Gamma=5/3$) polytropic equation of
state and the black hole as a Newtonian point mass which accretes
matter via an absorbing boundary at the Schwarzschild radius. Our
initial conditions correspond to tidally locked binaries in
equilibrium. The dynamical evolution is followed for approximately
23~ms, and in every case for $\Gamma=5/3$ we find that the neutron
star is tidally disrupted on a dynamical timescale, forming a dense
torus around the black hole that contains a few tenths of a solar
mass. A nearly baryon--free axis is present in the system throughout
the coalesence, a fireball expanding along that axis would avoid
excessive baryon contamination and could give rise to a modestly
beamed gamma--ray burst.
\end{abstract}

\section{Introduction}\label{intro}

Theoretical studies of the binary coalescence of a neutron star with a
black hole are as venerable as the Texas Symposia
(Wheeler~1971~\cite{wheel}). Such coalescences have been implicated in
the creation of heavy nuclei through the r--process and were suspected
of causing the violent outbursts known as gamma--ray bursts (Lattimer
\& Schramm~1974, 1976~\cite{latt74,latt76}; they are also candidate
sources for the gravitational radiation detectors currently coming of
age, as the double neutron star binaries have been for a long time
(Clark \& Eardley 1977~\cite{clark}).

No binary system composed of a black hole and a neutron star has as
yet been identified as such by the astronomers. However, theoretical
estimates and studies of binary stellar evolution predict that such
systems are created, and that the expected rate of their coalescence
may be between about one per one hundred thousand and one per a
million years per galaxy (Lattimer \& Schramm~1974,
1976~\cite{latt74,latt76}; Narayan, Piran \& Shemi 1991~\cite{nara};
Tutukov \& Yungelson 1993~\cite{tutukov}; Lipunov, Postnov \&
Prokhorov 1997~\cite{lipunov}). This rate is about right to explain
the r--process and gamma--ray bursts (although it remains to be
demonstrated that the phenomena are related to the coalescence), and
would give a very satisfactory event rate for future gravity wave
detectors (10 to 100 per year out to 200 Mpc).

The expected outcome of the coalescence ranged from the neutron star
``plunging'' into the black hole (Bardeen, Press \& Teukolsky
1972~\cite{bardeen}), through the disruption of the neutron star and
the formation of a transient accretion stream (Wheeler
1971~\cite{wheel}, Lattimer \& Schramm 1976~\cite{latt76}) or of an
accretion disk/torus (Paczy\'nski 1991~\cite{bp91}, Jaroszy\'nski
1993~\cite{jaro}, Witt {\it et al.} 1994~\cite{witt}), to rapid growth
of the binary separation in a steady Roche-lobe overflow scenario
(Blinnikov {\it et al.}  1984~\cite{blinnikov}, Portegies Zwart
1998~\cite{port}).  It seemed important to investigate the
hydrodynamics of the process numerically, to determine which, if any,
of these outcomes are likely.

We have initially performed Newtonian simulations treating the neutron
star as a stiff polytrope (Lee \& Klu\'zniak 1995~\cite{acta}),
motivated by a desire to determine the timescale of the coalescence
with the black hole and to investigate the spatial distribution of the
matter lost by the star. These questions were of relevance to the
theory of gamma--ray bursts, which seems to allow the coalescence to
power the burst in the accepted relativistic shock model, provided
that the environment into which the fireball expands has a very small
number of baryons at least in some directions (M\'esz\'aros \& Rees
1992, 1993~\cite{rees92,rees93}), and that the explosive event is
highly variable in time (Sari \& Piran 1997~\cite{sari}). We found the
coalescence of a neutron star and a black hole promising in these
respects, at least in our Newtonian model (Klu\'zniak \& Lee
1998~\cite{kl}).

However, the stiff polytrope ($\Gamma=3$), used by us in the work
cited above, differed in one crucial respect from a neutron star---it
responded to mass loss by shrinking, instead of expanding. For this
reason we have repeated the simulation with a softer polytrope, and we
report the results below, and elsewhere (Lee \& Klu\'zniak
1999~\cite{LKII}).

\section{Numerical method}\label{method}

For the simulations presented here, we have used the method known as
Smooth Particle Hydrodynamics. Our code is three--dimensional and is
completely Newtonian, although removal of angular momentum by
gravitational radiation was included, as described below. This method
has been described often, we refer the reader to
Monaghan~(1992)~\cite{monaghan} for a review of the principles of SPH,
and to Lee~(1998)~\cite{phd} and Lee \& Klu\'{z}niak~(1998)~\cite{LKI}
for a detailed description of our own code.

Here, we model the neutron star via a polytropic equation of state,
$P=K \rho^{\Gamma}$ with $\Gamma=5/3$. The unperturbed (spherical)
neutron star has a radius R=13.4~km and mass M=1.4M$_{\odot}$. The
black hole (of mass $M_{\rm BH}$) is modeled as a Newtonian point
mass, with a potential \( \Phi_{\rm BH}(r) = -GM_{\rm BH}/r \). We
model accretion onto the black hole by placing an absorbing boundary
at the Schwarzschild radius ($r_{Sch}=2GM_{\rm BH}/c^{2}$). Any
particle that crosses this boundary is absorbed by the black hole and
removed from the simulation. The mass and position of the black hole
are continously adjusted so as to conserve total mass and total
momentum.

Initial conditions corresponding to tidally locked binaries in
equilibrium are constructed in the co--rotating frame of the binary
for a range of separations {\em r} and a given value of the mass ratio
$q=M_{\rm NS}/M_{\rm BH}$~(Rasio \& Shapiro~1994~\cite{RS94}; Lee \&
Kluzniak~1998~\cite{LKI}). The neutron star is modeled with $N=17,256$
particles at the start of the calculation in every case presented
here. We have calculated the gravitational radiation signal emitted
during the coalescence in the quadrupole approximation, and refer the
reader to Lee \& Klu\'{z}niak~(1999)~\cite{LKII} for details.

We have included a term in the equations of motion that simulates the
effect of gravitational radiation reaction on the components of the
binary system in the quadrupole approximation (see Landau \&
Lifshitz~1975~\cite{landau}). This formulation of the gravitational
radiation reaction has been used in SPH before~(Davies et
al.~1994~\cite{davies}, Zhuge et al.~1996~\cite{zhuge}, Rosswog et
al.~1999~\cite{rosswog}) in the case of merging neutron stars, and it
is usually switched off once the stars come into contact, when the
point--mass approximation clearly breaks down. We are assuming then,
that the polytrope representing the neutron star can be considered as
a point mass for the purposes of including radiation
reaction. Clearly, the validity of this assumption must be verified
{\it a posteriori} when the simulation has run its course. If the
neutron star is disrupted during the encounter with the black hole,
this radiation reaction must be turned off, since our formula would no
longer give meaningful results. We have adopted a switch for this
purpose, as follows: if the center of mass of the SPH particles comes
within a prescribed distance of the black hole (effectively a tidal
disruption radius), then the radiation reaction is turned off. This
distance is set to $r_{tidal} = C R(M_{\rm BH}/M_{\rm NS})^{1/3}$,
where {\em C} is a constant of order unity.

\section{Results}\label{results}

\subsection{Evolution of the binary}\label{initial}

\begin{figure*}
\psfig{width=\textwidth,file=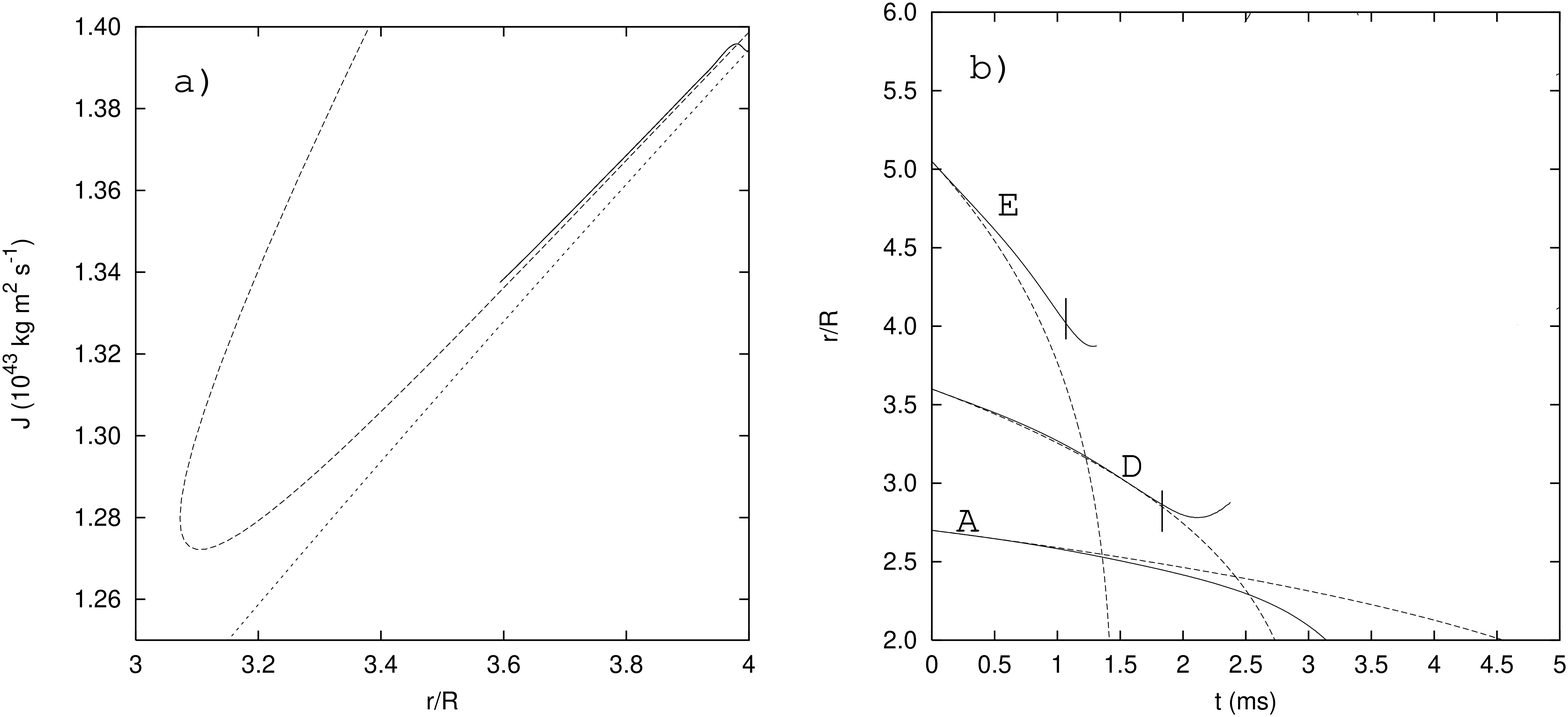,angle=0,clip=}
\caption{(a) Total angular momentum {\em J} as a function of binary
separation {\em r} along the equilibrium sequence for a mass ratio
$q$=0.31 (run D). The solid line is the result of the SPH calculation,
the dashed line results from approximating the neutron star as a
compressible tri--axial ellipsoid and the dotted line from
approximating it as a rigid sphere.(b) Binary separation during the
dynamical runs for various initial mass ratios (see
Table~\ref{parameters}). The vertical line across each curve (if
present), indicates the time when gravitational radiation reaction was
switched off.}
\label{jvsr}
\end{figure*}

To allow comparisons of results for differing equations of state, we
have run simulations with the same initial binary mass ratios as
previously explored for $\Gamma=3$~(Lee \&
Klu\'{z}niak~1998~\cite{LKI}), namely $q$=1, $q$=0.8 and
$q$=0.31. Additionally we have examined the case with mass ratio
$q$=0.1. Equilibrium sequences of tidally locked binaries were
constructed for a range of initial separations, terminating at the
point where the neutron star overflows its Roche Lobe (at
$r=r_{RL}$). In Figure~\ref{jvsr}a
we show the variation of total angular momentum {\em J} for one of
these sequences as a function of binary separation (solid
line). Following Lai, Rasio \& Shapiro~(1993b)~\cite{LRSb}, we have
also plotted the variation in {\em J} that results from approximating
the neutron star as compressible tri--axial ellipsoid (dashed lines)
and as a rigid sphere (dotted lines). In all cases, the SPH results
are very close to the ellipsoidal approximation until the point of
Roche--Lobe overflow. This result is easy to understand if one
considers that the softer the equation of state, the more centrally
condensed the neutron star is and the less susceptible to tidal
deformations arising from the presence of the black hole.

For $\Gamma=3$~(Lee \& Klu\'{z}niak~1998~\cite{LKI}), the variation in
angular momentum as a function of binary separation was qualitatively
different (for high mass ratios) from our present findings for
$\Gamma=5/3$. For $q$=1 and $q$=0.8, total angular momentum attained a
minimum at some critical separation {\em before } Roche--Lobe overflow
occurred. This minimum indicated the presence of a dynamical
instability, which made the binary decay on an orbital timescale. This
purely Newtonian effect arose from the tidal interactions in the
system~(Lai, Rasio \& Shapiro~1993a~\cite{LRSa}). In the present
study, we expect all orbits with initial separations $r\ge r_{RL}$ to
be dynamically stable.

For polytropes, the mass--radius relationship is \( R \propto
M^{(\Gamma-2)/(3\Gamma-4)}\). For $\Gamma$=5/3, this becomes \( R
\propto M^{-1/3}\). Thus, the polytrope considered here responds to
mass loss by expanding, as do neutron stars modeled with realistic
equations of state~(Arnett \& Bowers~1977~\cite{arnett})--the
dynamical disruption of the star reported below seems to be related to
this effect. For the polytropic index considered in Lee \&
Klu\'{z}niak~(1998)~\cite{LKI}, the star was not disrupted (see also
Lee \& Klu\'{z}niak 1995~\cite{acta}; 1997~\cite{hunt2}; Klu\'{z}niak
\& Lee 1998~\cite{kl}), but we find no evidence in any of our
dynamical calculations for a steady mass transfer in the binary, such
as the one suggested in the literature (e.g. Blinnikov et
al.~1984~\cite{blinnikov}; Portegies Zwart~1998~\cite{port}).

Using the quadrupole approximation, one can compute the binary
separation as a function of time for a point--mass binary, and obtain
\begin{eqnarray}
r=r_{i}\left( 1-t/t_{0} \right)^{1/4}, \label{eq:ptdecay}
\end{eqnarray}
with \( t_{0}^{-1}=256 G^{3}M_{\rm BH} M_{\rm NS} (M_{\rm BH}+M_{\rm
NS})/(5r_{i}^{4}c^{5})\). Here $r_{i}$ is the separation at $t$=0. For
black hole--neutron star binaries studied here, the timescale for
orbital decay because of angular momentum loss to gravitational
radiation, $t_{0}$, is on the order of the orbital period, $P$ (for
$q$=1, at an initial separation $r_{i}$=2.7R we find $t_{0}$=6.5~ms
and $P$=2.24~ms).

\subsection{Run parameters}

In Table~\ref{parameters} we present the parameters distinguishing
each dynamical run we performed. All times are in milliseconds and all
distances in kilometers. The runs are labeled with decreasing mass
ratio (increasing black hole mass), from $q$=1 down to $q$=0.1. All
simulations were run for the same length of time, $t_{final}=22.9$~ms
(this covers on the order of ten initial orbital periods for the mass
ratios considered). The initial separation for each dynamical run is
given as $r_{i}$, and the separation at which Roche Lobe overflow from
the neutron star onto the black hole occurs is given by $r_{RL}$.

\begin{table}
 \caption{Important parameters for each run}
 \label{parameters}
 \begin{tabular}{@{}ccccccc}
  Run & $q$ & $r_{RL}$(km) & $r_{i}$(km)  
        & $t_{rad}$(ms)
        & $t_{f}$(ms) & $N$ \\
  A   & 1.00 & 35.78 & 36.18 & 3.49 & 22.9 
        & 17,256 \\
  B   & 0.80 & 37.65 & 38.19 & 3.43 & 22.9 
        & 17,256 \\
  D   & 0.31 & 48.11 & 48.24 & 1.83 & 22.9 
        & 17,256 \\
  E   & 0.10 & 67.13 & 67.67 & 1.09 & 22.9 
        & 17,256 \\

 \end{tabular}

 \medskip
\end{table}

The fifth column in Table~\ref{parameters} shows the value of
$t_{rad}$, when radiation reaction is switched off according to the
criterion established in section~\ref{method}. We note here that run E
is probably at the limit of what should be inferred from a Newtonian
treatment of such a binary system. The black hole is very large
compared to the neutron star, and the initial separation
($r_{i}=67.87$~km) is such that the neutron star is well within the
innermost stable circular orbit around a Schwarzschild black hole of
the mass considered.

\subsection{Morphology of the mergers}\label{morphology}

The initial configurations are close to Roche Lobe overflow, and mass
transfer from the neutron star onto the black hole starts within one
orbital period for all runs presented here. In every run the binary
separation (solid lines in in Figure~\ref{jvsr}) initially decreases
due to gravitational radiation reaction. For high mass ratios, (runs
A, B) the separation decays faster than what would be expected of a
point--mass binary. This is also the case for a stiff equation of
state, in black hole--neutron star mergers~(Lee \&
Kluzniak~1998~\cite{LKI}) as well as in binary neutron star
mergers~(Rasio \& Shapiro~1994~\cite{RS94}), and merely reflects the
fact that hydrodynamical effects are playing an important role. For
the soft equation of state studied here, there is the added effect of
`runaway' mass transfer because of the mass--radius relationship (see
section~\ref{initial}). For run C, the solid and dashed lines in
Figure~\ref{jvsr}b follow each other very closely, indicating that the
orbital decay is primarily driven by angular momentum losses to
gravitational radiation. For run E, the orbit decays more slowly than
what one would expect for a point--mass binary. This is explained by
the fact that there is a large amount of mass transfer (10\% of the
initial neutron star mass has been accreted by $t=t_{rad}$ in this
case) in the very early stages of the simulation, substantially
altering the mass ratio in the system (the dashed curves in
Figure~\ref{jvsr}b are computed for fixed masses; at constant total
mass, note that from equation~\ref{eq:ptdecay}, lowering the mass
ratio in the system slows the orbital decay for $q<0.5$).

\begin{figure*}
\psfig{width=\textwidth,file=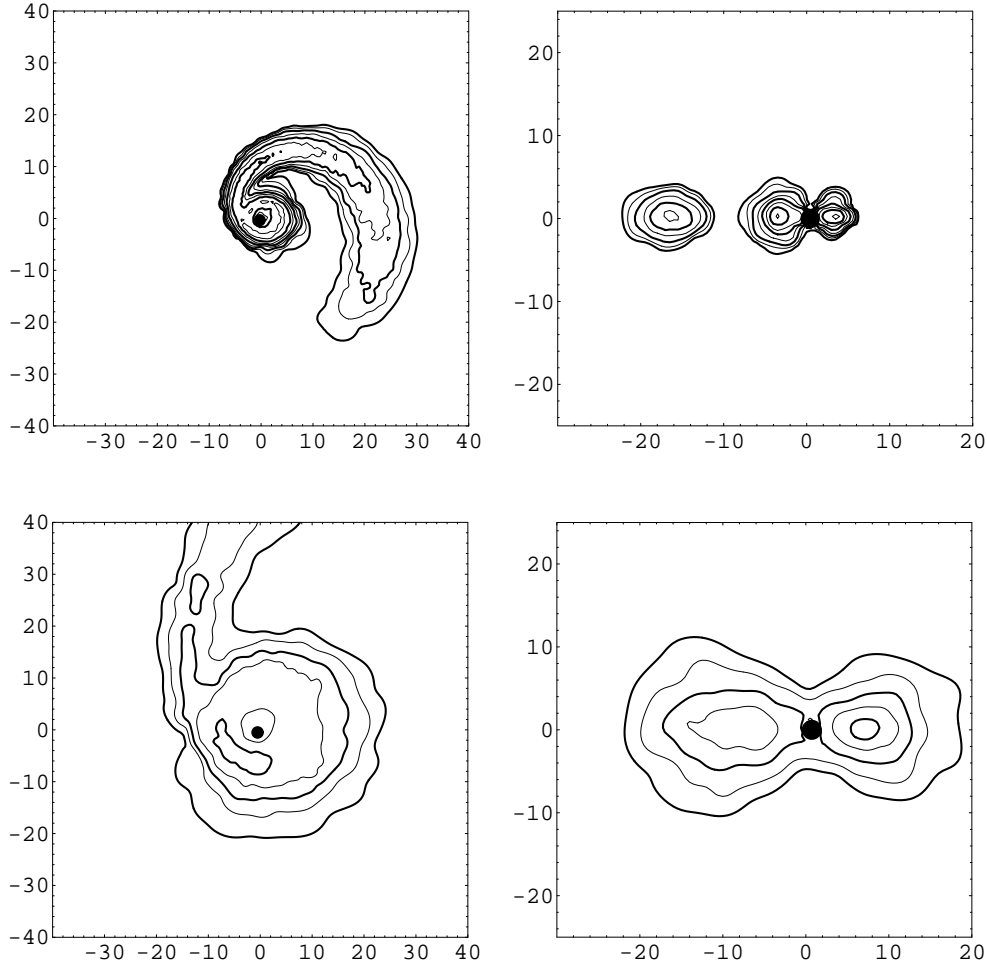,angle=0,clip=}
\caption{Density contours after the disruption of the polytrope for
$q=0.31$ (Run D) in the equatorial plane (left panels), and in the
meridional plane containing the black hole (right panels) at
$t$=5.73~ms (top panels) and $t$=22.9~ms (bottom panels). The axes are
labeled in units of the initial (unperturbed) neutron star radius
R=13.4~km. All contours are logarithmic and equally spaced every 0.5
dex. The lowest contour is at $\log{\rho/\rho_{0}}=-6$
($\rho_{0}=1.14\times10^{15}$g~cm$^{-3}$), and bold contours are
plotted at $\log{\rho/\rho_{0}}=-6,-5,-4,-3$.}
\label{rhocontoursq031b}
\end{figure*}

The general behavior of the system is qualitatively similar for every
run. Figure~\ref{rhocontoursq031b} shows density contours in the
orbital plane (left panels) and in the meridional plane containing the
black hole (right panels) for run D at $t=5.73$~ms and
$t=t_{f}=22.9$~ms.
\begin{figure*}
\psfig{width=\textwidth,file=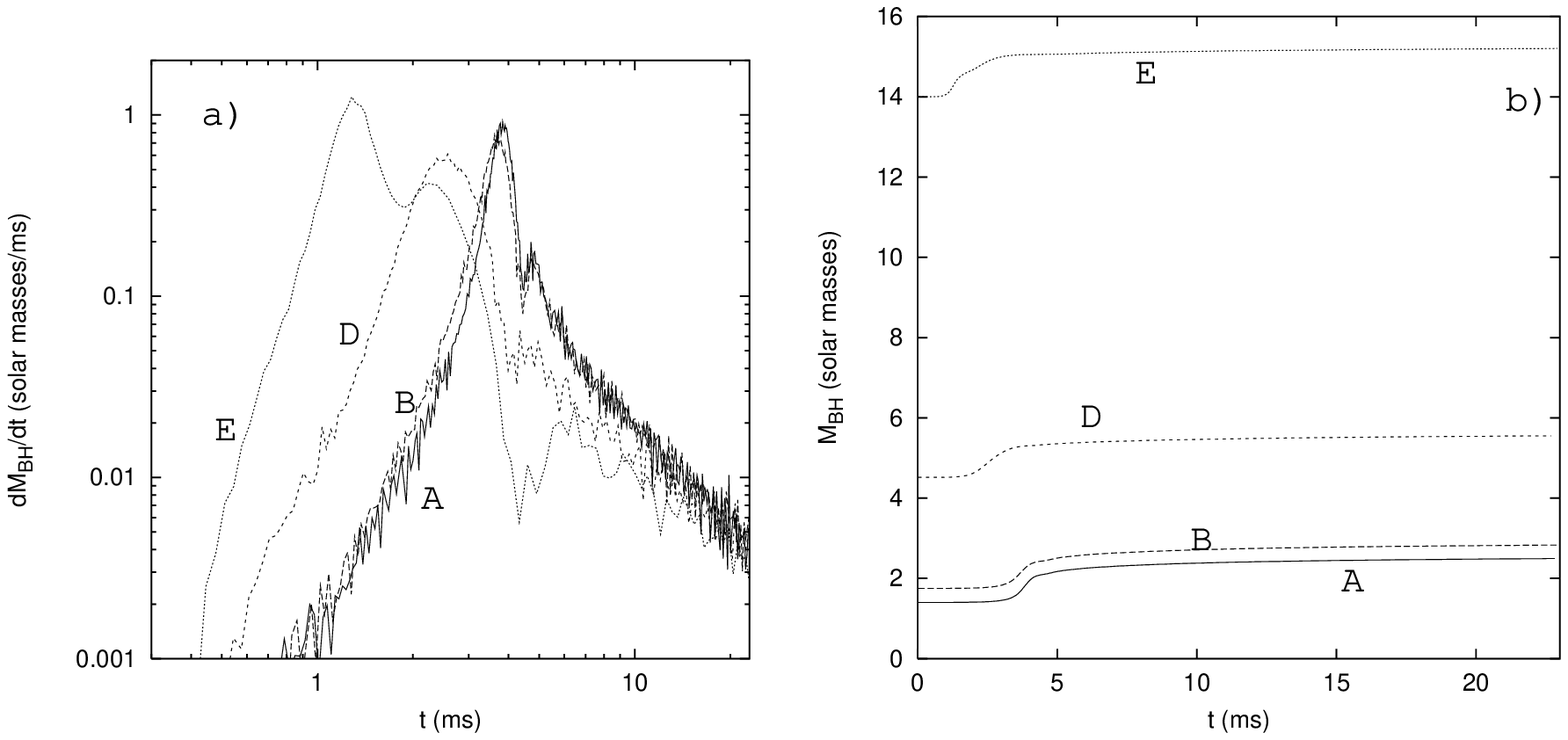,angle=0,clip=}
\caption{Mass accretion rate onto the black hole~(a) and mass of the
black hole~(b) for runs A, B, D and E.}
\label{mdot}
\end{figure*}
The neutron star becomes initially elongated along the binary axis and
an accretion stream forms, transferring mass to the black hole through
the inner Lagrange point. The neutron star responds to mass loss and
tidal forces by expanding, and is tidally disrupted. An accretion
torus forms around the black hole as the initial accretion stream
winds around it. A long tidal tail is formed as the material furthest
from the black hole is stripped from the star. Most of the mass
transfer occurs in the first two orbital periods and peak accretion
rates reach values between 0.5 M$_{\odot}$/ms and 1.2M$_{\odot}$/ms
(see Figure~\ref{mdot}). The mass accretion rate then drops and the
disk becomes more and more azimuthally symmetric, reaching a
quasi--steady state by the end of the simulations.

\begin{figure*}
\psfig{width=\textwidth,file=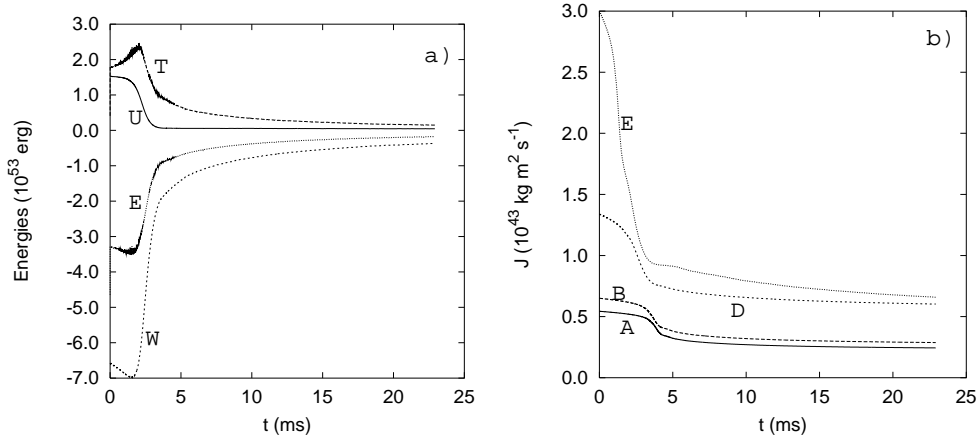,angle=0,clip=}
\caption{(a) Various energies of the system as a function of time for
run D. The kinetic (T), internal (U), gravitational potential (W) and
total (E) energies are indicated. (b) Angular momentum as a function
of time for every run.}
\label{energies}
\end{figure*}

We show in Figure~\ref{energies} the various energies of the system
(kinetic, internal, gravitational potential and total) for run D. 
The dramatic drop in total internal energy reflects the intense mass
accretion that takes place within the first couple of
orbits. Figure~\ref{energies} also shows [panel (b)] the total angular
momentum of the system for runs A, B, D and E (the only contribution
to the total angular momentum not plotted is the spin angular momentum
of the black hole, see below). Angular momentum decreases for two
reasons. First, if gravitational radiation reaction is still acting on
the system, it will decrease approximately according to the quadrupole
formula. Second, whenever matter is accreted by the black hole, the
corresponding angular momentum is removed from our system. In reality,
the angular momentum of the accreted fluid would increase the spin of
the black hole. We keep track of this accreted angular momentum and
exhibit its value in Table~\ref{GRB} as the Kerr parameter of the
black hole. This shows up as a decrease in the total value of {\em J}.

\begin{table*}
 \caption{Accretion disk structure}
 \label{disks}
 \begin{tabular}{@{}ccccccc}
  & & & & & & \\
  Run & $q$ & M$_{disk}$(M$_{\odot}$) & M$_{acc}$(M$_{\odot}$) 
        & $\dot{\rm M}_{max}$(M$_{\odot}$/ms)    
	& $\dot{\rm M}_{final}$(M$_{\odot}$/s)
        & $\tau_{disk}$(ms) \\
  A   & 1.00 & 0.263 & 1.092 & 0.831 & 4.88 & 54.1
        \\
  B   & 0.80 & 0.277 & 1.078 & 0.733 & 6.11 & 46.9
        \\
  D   & 0.31 & 0.316 & 1.036 & 0.549 & 4.88 & 63.1
        \\
  E   & 0.10 & 0.101 & 1.204 & 1.160 & 2.44 & 46.9
        \\

 \end{tabular}

 \medskip

\end{table*}

\begin{table*}
 \caption{}
 \label{GRB}
 \begin{tabular}{@{}ccccc}
 & & & & \\
  Run & $J_{\rm BH}c/G M_{\rm BH}^{2}$ &
        $\theta_{-3}$
        & $\theta_{-4}$ & $\theta_{-5}$ \\
  A   & 0.517 & 20 & 12 & 8 \\
  B   & 0.497 & 25 & 10 & 3 \\
  D   & 0.173 & 40 & 21 & 10 \\
  E   & 0.114 & 52 & 42 & 32 \\

 \end{tabular}

 \medskip

$\theta_{-n}$ is the half--angle (in degrees) of a cone above the
black hole and along the rotation axis of the binary that contains a
mass $M=1.4 \times 10^{-n}$M$_{\odot}$.

\end{table*}

\begin{figure*}
\psfig{width=\textwidth,file=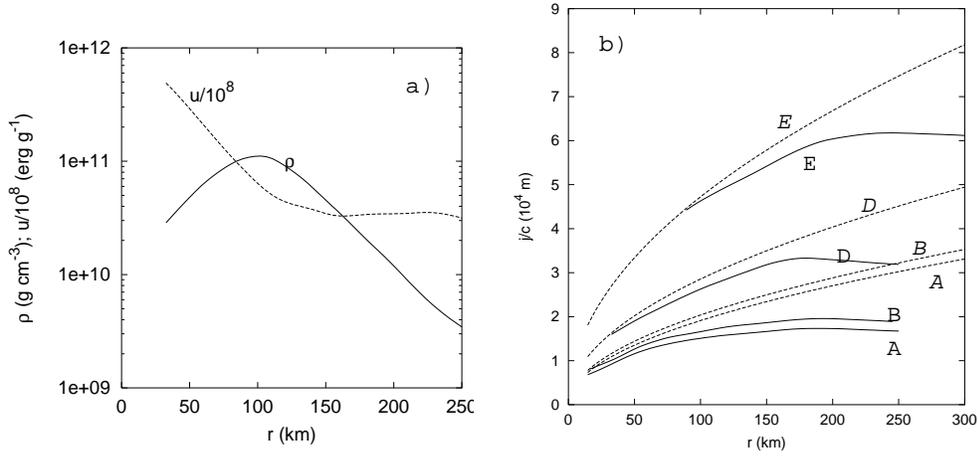,angle=0,clip=}
\caption{(a) Azimuthally averaged profiles for the density, $\rho$,
and the specific internal energy, {\em u} ($u/10^{8}$ is plotted), of
the accretion torus at $t=t_{f}$ for run D. The inner edge of the
curves is at $r=2 r_{Sch}$. At this stage in the simulation the torus
is close to being azimuthally symmetric. (b) Distribution of specific
angular momentum {\em j} for runs A, B, D and E at $t=t_{f}$ (solid
lines, A, B, D, E) and the specific angular momentum of a Keplerian
accretion disk for the same black hole mass (dashed lines, {\it A},
{\it B}, {\it D}, {\it E}).}
\label{diskprofiles}
\end{figure*}

\subsection{Accretion disk structure}

In Table~\ref{disks} we show several parameters pertaining to the
final accretion structure around the black hole for every run. The
mass that has been accreted by the black hole is denoted by
M$_{acc}$. The disk settles down to a fairly azimuthally symmetric
structure within a few initial orbital periods (except for the long
tidal tail, which always persists as a well--defined structure), and
there is a baryon--free axis above and below the black hole in every
case (see below). We have calculated the mass of the remnant disk,
$M_{disk}$, by searching for the amount of matter that has sufficient
specific angular momentum {\em j} at the end of the simulation to
remain in orbit around the black hole (as in Ruffert \&
Janka~1998~\cite{ruffert98}). This material has
$j>j_{crit}=\sqrt{6}GM_{t}/c$, where $M_{t}$ is the total mass of the
system. By the end of the simulations, between 70\% and 80\% of the
neutron star has been accreted by the black hole.  It is interesting
to note that the {\em final} accretion rate (at $t=t_{f}$) appears to
be rather insensitive to the initial mass ratio, and is between
2.4~M$_{\odot}$~s$^{-1}$ and 6.1~M$_{\odot}$~s$^{-1}$. From this final
accretion rate we have estimated a typical timescale for the evolution
of the accretion disk, $\tau_{disk}=M_{disk}/\dot{M}_{final}$. Despite
the difference in the initial mass ratios and the typical sizes of the
disks, the similar disk masses and final accretion rates make the
lifetimes comparable for every run.

We have plotted azimuthally averaged density and internal energy
profiles in Figure~\ref{diskprofiles} for run D. The specific internal
energy is greater towards the center of the disk, and flattens out at
a distance from the black hole roughly corresponding the density
maximum, at $u\simeq 3 \times 10^{18}$~erg~g$^{-1}$, or
3.1~MeV/nucleon, and is largely independent of the initial mass
ratio. The inner regions of the disks have specific internal energies
that are greater by approximately one order of magnitude.

Additionally, panel (b) in the same figure shows the azimuthally
averaged distribution of specific angular momentum {\em j} in the
orbital plane for all runs. The curves terminate at $r_{in}=2r_{Sch}$.
Pressure support in the inner regions of the accretion disks makes the
rotation curves sub--Keplerian, while the flattening of distribution
marks the outer edge of the disk and the presence of the long tidal
tail (see Figure~\ref{rhocontoursq031b}),which has practically
constant specific angular momentum.

The Kerr parameter of the black hole, given by $a=J_{\rm BH}c/G M_{\rm
BH}^{2}$, is also shown in Table~\ref{GRB}. We have calculated it from
the amount of angular momentum lost by the fluid via accretion onto
the black hole (see Figure~\ref{energies}b), assuming that the black
hole is {\em not} rotating at $t=0$. The specific angular momentum of
the black hole is smaller for lower mass ratios simply because the
black hole is initially more massive when {\em q} is smaller.

It is of crucial importance for the production of GRBs from such a
coalescence event that there be a baryon--free axis in the system
along which a fireball may expand with ultrarelativistic
velocities~(M\'{e}sz\'{a}ros \& Rees 1992,
1993~\cite{rees92,rees93}). We have calculated the baryon
contamination for every run as a function of the half--angle $\Delta
\theta$ of a cone directly above the black hole and along the rotation
axis of the binary that contains a given amount of mass $\Delta
M$. Table~\ref{GRB} shows these angles (in degrees) for $\Delta M/{\rm
M}_{\odot}=1.4\times 10^{-3},1.4\times 10^{-4}, 1.4\times
10^{-5}$. There is a greater amount of pollution for high mass ratios
(the disk is geometrically thicker compared to the size of the black
hole), but in all cases only modest angles of collimation are required
to avoid contamination. We note here that the values for $\theta_{-5}$
are rough estimates at this stage since they are at the limit of our
numerical resolution in the region directly above the black hole.

\section{Discussion}

The numerical simulations reported here were quasi-newtonian.  An
important caveat to keep in mind is that inclusion of general
relativistic effects may lead to results qualitatively different from
even post-newtonian treatment (Wilson, Mathews \& Marronetti
1996~\cite{wilson96}).

Our results indicate that the outcome of the binary coalescence
depends on the nature of the star orbiting the black hole. As reported
previously (Lee \& Klu\'zniak 1995~\cite{acta}, 1998~\cite{LKI},
Klu\'zniak \& Lee 1998~\cite{kl}), when we modeled the star as a
polytrope with adiabatic index $\Gamma=3$, the coalescence appeared to
be an intermittent process in which the core of the polytrope survives
the initial encounters and increases its separation from the black
hole, thus extending the merger to possibly $\sim 0.1$~s.  For the
softer polytrope discussed here ($\Gamma=5/3$), the star is disrupted
completely in a few milliseconds and all that remains after the
initial mass transfer is an accretion disk, containing no more than
$1/5$ of the initial mass, and some ejecta.  Perhaps the current
simulation with $\Gamma=5/3$, is the more realistic one, because for
this polytrope $dM/dR <0$, as for physical models of neutron stars
(e.g. Arnett \& Bowers 1977~\cite{arnett}).

In agreement with earlier suggestions (Lattimer \& Schramm~1974,
1976~\cite{latt74,latt76}), we have found that some matter will be
ejected from the system, in an amount sufficient to account for the
abundance of the r--process nuclei (assuming the r--process does
indeed occur during the merger).

The binary coalescence of a neutron star with a black hole remains an
attractive theoretical source of gamma--ray bursts. The energy
requirements for at least one recently observed burst are so severe,
if emission is isotropic (e.g. Kulkarni {\it et al.}
1998~\cite{kulkarni}), that some degree of beaming seems
desirable. According to the simulations presented here,
ultrarelativistic flows are possible in the post-merger system only
along the rotational axis of the system in a solid angle of about
$0.1$~steradian. Proper inclusion of neutrino transport may change
this angle somewhat.  The rather short ($\sim 50$~ms) accretion
timescale of the remnant disk reported, does not include possible
interaction between the disk and the black hole (Blandford \& Znajek
1977~\cite{blandford}). In fact, the appearance in the simulation of a
substantial toroidal disk around the black hole is encouraging, as it
may allow the black hole spin to be extracted by the Blandford-Znajek
mechanism, possibly powering in this manner the gamma--ray burst
fireball (M\'{e}sz\'{a}ros \& Rees 1997~\cite{rees97}).

\section*{ACKNOWLEDGMENTS}

We gratefully acknowledge support for this work from DGAPA--UNAM and
KBN (grant P03D01311).


\section*{References}

\begin{thebibliography}{99}

\bibitem{wheel} Wheeler, J.A., 1971, Pontificae Acad. Sci. Scripta
Varia 35, 539

\bibitem{latt74} Lattimer, J.M. \& Schramm, D.N., 1974, ApJ 192, L145

\bibitem{latt76} Lattimer, J.M., Schramm, D.N., 1976, ApJ 210, 549

\bibitem{clark} Clark, J.P.A. \& Eardley, D.M., 1977, ApJ 215, 311

\bibitem{nara} Narayan R., Piran T. \& Shemi A., 1991, ApJ, 379, L17

\bibitem{tutukov} Tutukov, A. V. \& Yungelson, L. R., 1993, MNRAS, 260,
675

\bibitem{lipunov} Lipunov, V.M., Postnov, K.A. \& Prokhorov, M.E., 1997
New Astronomy, vol. 2, 43

\bibitem{bardeen} Bardeen, J.M., Press, W.H. \& Teukolsky, S.A., 1972,
ApJ 178, 347

\bibitem{bp91} Paczy\'{n}ski, B., 1991, Acta Astron. 41, 257

\bibitem{jaro} Jaroszy\'{n}ski, M., 1973, Acta Astron. 43, 183

\bibitem{witt} Witt, H. J., Jaroszy\'{n}ski, M., Haensel, P.,
Paczy\'{n}ski, B. \& Wambsganss, J., 1994, ApJ 422, 219

\bibitem{blinnikov} Blinnikov S.I., Novikov I.D., Perevodchikova T.V.,
Polnarev A.G., 1984, PAZh, 10, 422 [SvAL 10, 177].

\bibitem{port} Portegies Zwart S.F., 1998, ApJ, 503, L53

\bibitem{acta} Lee, W.H., Klu\'{z}niak, W., 1995, Acta Astron. 45, 705

\bibitem{rees92} M\'{e}sz\'{a}ros, P., Rees, M.J., 1992, MNRAS 257, 29P

\bibitem{rees93} M\'{e}sz\'{a}ros, P., Rees, M.J., 1993, ApJ 405, 278

\bibitem{sari} Sari, R. \& Piran, T., 1997, ApJ 485, 270

\bibitem{kl} Klu\'{z}niak, W. Lee, W.H., 1998, ApJ 494, L53

\bibitem{LKII} Lee, W.H., Klu\'zniak, W., 1999, MNRAS submitted

\bibitem{monaghan} Monaghan, J.J., 1992, ARA\&A, 30, 543

\bibitem{phd} Lee, W.H., 1998, Ph.D. Thesis, University of Wisconsin

\bibitem{LKI} Lee, W.H., Klu\'{z}niak, W., 1998, ApJ submitted,
astro-ph/9808185

\bibitem{RS94} Rasio, F., Shapiro, S.L. 1994, ApJ 432, 242

\bibitem{landau} Landau L.D., Lifshitz E.M., 1975, The Classical
Theory of Fields, Heinemann, Oxford.

\bibitem{davies} Davies M.B., Benz W., Piran T., Thielemann F.K.,
1994, ApJ, 431, 742

\bibitem{zhuge} Zhuge X., Centrella J.M., McMillan S.L.W., 1996,
Phys. Rev. D, 54, 7261

\bibitem{rosswog} Rosswog S., Liebend\"{o}rfer M., Thielemann F.K.,
Davies M.B., Benz W., Piran T., 1999, A\&A, 341, 499

\bibitem{LRSb} Lai, D., Rasio, F., Shapiro, S.L., 1993b, ApJS 88, 205

\bibitem{LRSa} Lai, D., Rasio, F., Shapiro, S.L., 1993a, ApJ 406, L63

\bibitem{arnett} Arnett W.D., Bowers R.L., 1977, ApJS, 33, 415

\bibitem{hunt2} Lee W.H., Klu\'{z}niak W., 1997, in Meegan C., Preece
R., Koshut P. eds., AIP Proc. 428, Gamma Ray Bursts, AIP, New York,
p.~798

\bibitem{ruffert98} Ruffert M., Janka H.-Th., 1999, A\&A, 344, 573

\bibitem{wilson96} Wilson, J.R., Mathews, G.J. \& Marronetti, P.,
1996, Phys. Rev. D 54, 1317

\bibitem{kulkarni} Kulkarni, S.R. {\it et al.}, 1998, Nature 393, 215

\bibitem{blandford} Blandford, R.D., Znajek, R.L., 1977, MNRAS 179,
433

\bibitem{rees97} M\'{e}sz\'{a}ros, P. \& Rees, M.J., 1997, ApJ 482, L29


\end{thebibliography}
\end{document}